\documentstyle[epsfig,prl,twocolumn,aps]{revtex}
\tighten
\begin{document}
\draft
\title{The effect of the spin-orbit geometric phase on the spectrum of 
Aharonov-Bohm oscillations in a semiconductor mesoscopic ring}
\author{A. G. Mal`shukov and V. V. Shlyapin}
\address{Institute of Spectroscopy, Russian Academy of Science,
142092 Troitsk, Moscow oblast, Russia}
\author{K. A. Chao}
\address{Department of Theoretical Physics, Lund University,
S-223 62 Lund, Sweden}
\date{\today}
\maketitle

\begin{abstract}
Taking into account the spin precession caused by the spin-orbit splitting
of the conduction band in semiconductor quantum wells, we have calculated
the Fourier spectra of conductance and state-density correlators in a 2D 
ring, in order to investigate the structure of the main peak corresponding
to Aharonov-Bohm oscillations. In narrow rings the peak structure is
determined by the competition between the spin-orbit and the Zeeman
couplings. The latter leads to a peak broadening, and produces the peak
splitting in the state-density Fourier spectrum. We have found an
oscillation of the peak intensity as a function of the spin-orbit coupling
constant, and this effect of the quantum interference caused by the spin
geometric phase is destroyed with increasing Zeeman coupling.
\end{abstract}


The spin-orbit interaction (SOI) gives rise to a geometric phase in the
quantum amplitude of a particle propagating along a closed trajectory.
In ideal 1D rings this can lead to quantum oscillations of transport
parameters similar to the Aharonov-Bohm effect~\cite{al,yqs}. In
disordered conductors the interference between two time reversed paths
produces the oscillation of mean conductance analogous to the 
Altshuler-Aronov-Spivak effect. Such oscillation in 1-D systems was shown
by Meir, Gefen and Entin-Wohlman~\cite{mge}, and in systems of higher
dimensions by Mathur and Stone~\cite{ms}. Besides, SOI also modifies the
shape of the Aharonov-Bohm and Altshuler-Aronov-Spivak oscillations. For
a disordered material, the mean conductance is obtained as an ensemble
average over a large number of measurements on different samples. One can
try to detect the quantum effects associated to the spin-orbit phase by
measuring the oscillations of mean conductance when the spin-orbit
coupling strength or the external magnetic field is varied. To our
knowledge, such experiments have not yet established any evidence of the
spin-orbit geometric phase.

For a disordered material, if one takes the Fourier transform of the mean
conductance $\langle g(B)\rangle$ as a function of the external magnetic
field $B$, the spectrum is dominated by the Altshuler-Aronov-Spivak
oscillations which is periodic in magnetic flux with a period $hc/2e$. On
the other hand, if one takes first the Fourier transform $g(\nu)$ of a
measured conductance $g(B)$, and then performs an ensemble average $\langle
|g(\nu)|\rangle$ of the Fourier amplitude, one would expect that the
so-derived spectrum will exhibit a main peak corresponding to the
Aharonov-Bohm oscillations with a period $hc/e$ in magnetic flux~\cite{ww}.
Consequently, the average of Fourier amplitude,
$\langle |g(\nu)|\rangle$ represents correlations of conductances measured
at different magnetic fields. The dependence of these correlations on the
SOI can then manifest itself in the shape of the mean peak. In a recent
experiment~\cite{exp} on mesoscopic rings made from a AlSb/InAs quantum
well structure, the data were analyzed in this way for the first time, and
a split of the main peak in the measured spectrum $\langle |g(\nu)|\rangle$
was observed. The authors of Ref.~\onlinecite{exp} have conjectured that
the observed splitting is due to the strong Rashba SOI in a doped AlSb/InAs
quantum well.

While the experimental results in Ref.~\onlinecite{exp} remain to be
explained, in our opinion, the fundamental question needed to be answered
is how the Fourier spectrum of the conductance correlations, particularly
the shape of the main Fourier peak, is influenced by the interplay between
the spin-orbit phase and the external magnetic field. This is the aim of
the present Letter. The more suitable starting point for such a theoretical
analysis is
\begin{equation}\label{gnu}
\langle |g(\nu )|^{2}\rangle
= \int \int_{-B_{0}}^{B_{0}}dB\,dB^{\prime }\ e^{i\nu
(B-B^{\prime })}\langle g(B)g(B^{\prime })\rangle \, ,
\end{equation}
where the interval between $-B_{0}$ and $B_{0}$ covers the region in which
$g(B)$ is measured. This interval is much larger than both the period of
the Aharonov-Bohm oscillations and the magnetic field correlation
range~\cite{meso} of mesoscopic fluctuations. In this Letter we will
calculate (\ref{gnu}) in the diffusion regime of a disordered 2D
semiconductor ring of width $w$ and radius $R$. Our theory explains the
main physical mechanisms which determine the shape of the Fourier spectrum.
For the state-density correlator which partly contributes to (\ref{gnu})
we predict a split of the main Fourier peak when effect of the Zeeman
interaction is not completely suppressed by SOI and mesoscopic fluctuations.
After we point out that this same feature appears in an ideal ballistic
ring, we conjecture later that our theory also provides the physical origin
of main peak splitting observed in a chaotic ballistic ring~\cite{exp}.

Since it is the Rashba~\cite{rash} term rather than the
Dresselhaus~\cite{dress} term which gives the major contribution to the SOI
in InAs based QW~\cite{parsoi}, for simplicity in this Letter we will
neglect the Dresselhaus term. We will use the standard perturbation theory
which was applied previously~\cite{meso} to analyze the conductance
correlations. In the framework of this theory the correlator in (\ref{gnu})
is expressed via two-particle propagators (Cooperons and diffusons) where
one of the particles propagates at the magnetic field $B$ and the other at
$B^{\prime}$. If one neglects the Zeeman interaction with the external
magnetic field, the Cooperon propagator is a function of $B$+$B^{\prime}$,
while the diffuson depends only on $\Delta B$=$B$-$B^{\prime}$. Hence, in
the vicinity of the main peak at $\nu$=$2\pi^2R^2/\Phi_0$, where 
$\Phi_0$=$hc/e$ is the flux quantum, the major contribution to (\ref{gnu})
is given by diffusons. If the Zeeman interaction is taken into account, the
Cooperon propagator is no longer a function of $B$+$B^{\prime}$ alone.
Nevertheless, Cooperon's contribution is relatively small if the Zeeman
energy $g\mu_B\Phi_0/\pi R^2$ is much less than the Thouless energy
$E_T$=$D\hbar /R^2$. Therefore, in the integrand of (\ref{gnu}) we will
retain only the part of the correlator associated to diffusons.

Each of the diffusons is a component of a matrix ${\bf D(r,r}^{\prime})$
with four spin indices representing the spin states of an electron and a
hole. Following Ref.~\cite{meso} one can show that the correlator in the
integrand of (\ref{gnu}) is proportional to
\begin{eqnarray}\label{corr}
&& \int d^2r d^2r^{\prime} \{Re \left(
Tr[{\bf D(r,r}^{\prime}){\bf D}({\bf r}^{\prime},{\bf r})] \right)
\nonumber \\
&& + 2Tr[{\bf D(r,r}^{\prime}){\bf D}^{\dag}({\bf r},{\bf r}^{\prime})]\}
\, .
\end{eqnarray}
In the above equation the trace is taken separately over the electron and
hole spin indices. It is convenient to express ${\bf D}$ in the
representation of the total spin ${\bf S}$ of the electron-hole
two-particle system~\cite{alt}. In this representation ${\bf D}$ is a
4$\times$4 matrix with components $D_{mn}$. The indices $m$ and $n$ can
have the values -1,0,1 (for the $z$ component of the triplet) and $s$ (for
the singlet).

In A$_{3}$B$_{5}$ semiconductors the spin-orbit coupling has the form
$H_{so}$=${\bf h}_{{\bf k}}$$\cdot$${\bf s}$ for an electron having
spin ${\bf s}$ and quasi-momentum ${\bf k}$. This SOI and the
Zeeman interaction determine the spin dependence of ${\bf D}$, which can
be written as~\cite{soi}
\begin{eqnarray}
\tau && \langle (-i{\bf v}\cdot\nabla +
\frac{e}{c}{\bf v}\cdot\Delta {\bf A} +
{\bf h}_{{\bf k}}\cdot {\bf S)}^2 \rangle_{ang}
{\bf D(r,r}^{\prime}) \nonumber \\
&& + {\bf ZD(r,r}^{\prime}) +
\frac{1}{\tau _{\varphi}}{\bf D(r,r^{\prime})} 
= \delta ({\bf r}-{\bf r}^{\prime }) \, ,
\label{diff}
\end{eqnarray}
where $\tau _{\varphi}$ is the elastic mean free scattering time, $\tau$
the phase breaking time, and the notation $\langle\cdots\rangle_{ang}$
is an angular average over the Fermi surface. The term
{\bf ZD(r,r}$^{\prime}$) in the above equation is due to the Zeeman
interaction, and the nonzero components of {\bf Z} are
\begin{eqnarray}\label{zeeman}
Z_{0s} &=&Z_{s0}=\frac{ig\mu _{B}}{2} (B-B^{\prime })  \nonumber \\
Z_{11} &=&-Z_{-1-1}=\frac{ig\mu _{B}}{2} (B+B^{\prime }) \, .
\end{eqnarray}
We see that if $B$$\neq$$B^{\prime}$ the matrix {\bf Z} contains
components which mix the singlet part and the triplet part of the
diffuson.

We choose the gauge such that for the field difference $\Delta B$ the
vector potential is $\Delta {\bf A}$=$\Delta Br{\bf t}/2$, where {\bf t} is
a unit vector tangential to the ring. Since
$H_{so}$=$\alpha (k_{x}{\bf s_{y}}$-$k_{y}{\bf s_{x}})$ when the Rashba
term dominates, the boundary conditions at the inner and the outer radii of
the ring are
\begin{equation}\label{boundary}
-i\frac{\partial}{\partial r}{\bf D(r,r}^{\prime }) -
\frac{\alpha m^{*}}{\hbar}({\bf t}\cdot{\bf S}){\bf D(r,r}^{\prime})=0 \, .  
\end{equation}
In a narrow ring with the width $w$ much less than the radius $R$,
${\bf D(r,r}^{\prime})$ varies slowly across the annulus. If the
elastic mean free path $l$ is much shorter than $w$, such slow variation
can be treated perturbatively in the diffusion approximation~\cite{feng}.
Using the boundary conditions (\ref{boundary}), after averaging over $r$,
(\ref{diff}) is reduced to an effectively 1D equation, where {\bf D}
depends only on the azimuthal angles $\varphi,\varphi'$. This function
can be expressed in the form
\begin{equation}\label{expansion}
{\bf D}(\varphi,\varphi ^{\prime })=e^{i
{\bf S}_{z}(\varphi^{\prime}- \varphi)}\sum_n {\bf M}_{n}
(\varphi ^{\prime })e^{i\varphi n} \, .
\end{equation}
Making use of the rotation properties of the spin operator
\[
\exp [i{\bf S}_{z}(\varphi -\varphi ^{\prime })]({\bf S\cdot n)}
\exp [i{\bf S}_{z}(\varphi^{\prime } -\varphi)] =
({\bf S\cdot n}^{\prime }) \, ,
\]
where $n_x^{\prime}$=$\cos\varphi^{\prime}$ and
$n_y^{\prime}$=$\sin\varphi^{\prime}$, we arrive at a set of 4$\times$4
algebraic equations for the components of the matrices ${\bf M}_{n}$. From
these equations one can derive the following subset of equations which
contains only the -1,0,1 components of the matrices ${\bf M}_n$
\begin{eqnarray}\label{mdiff}
&& \left[n-\Delta \phi -\zeta ({\bf S\cdot N)}\right] ^{2}{\bf M}_n + 
[a + \rho^2 M_0 (1-{\bf S}^{2}_z)] (\Delta \phi)^2 {\bf M}_n \nonumber \\
&& + i \rho {\bf S}_z (\phi+\phi^{\prime}) {\bf M}_n +
\frac{1}{\tau _{\varphi }E_{T}}{\bf M}_n ={\bf 1} \, ,
\end{eqnarray}
where $a$=$w^2/4R^2$, 
$M_0$=$[(n$-$\Delta\phi)^2$+$a(\Delta\phi)^2$+$1/\tau_{\varphi}E_T]^{-1}$,
$\zeta^2$=1+$(R\alpha m^*/\hbar)^2$, and
$\rho$=$gm^*/k_flm$. {\bf N} is a unit vector with
$N_x$=$n_x^{\prime}(1$-$\zeta^{-2})^{1/2}$,
$N_y$=$n_y^{\prime}(1$-$\zeta^{-2})^{1/2}$ and $N_z$=$1/\zeta$.

The three dimensionless parameters $a$, $\zeta$ and $\rho$ determine the
shape of the main peak in $\langle |g(\nu)|^{2}\rangle$
given by (\ref{gnu}). $a$ describes the dephasing due to the penetration
of the magnetic flux $\Delta\phi$=$\Delta B\pi R^2/\Phi_0$ into the annulus
of the ring. We should remind the reader that to calculate $a$ we have
assumed diffusive propagation of particles in the radial direction. However, it
is reasonable to believe that the dependence on flux of the form
$a(\Delta\phi)^2$, as appearing in (\ref{mdiff}), is also valid for rings
with ballistic transport along radial direction. In this case $a$ can be
considered as a phenomenological parameter. The spin-orbit coupling
constant $\zeta$ gives the spin-phase winding number after a particle has
traversed a closed path along the ring. The parameter $\rho$ is related to
the Zeeman interaction. It determines the amount of mixing between the
triplet and the singlet components of the diffusion propagator. This
mixing appears in (\ref{mdiff}) in the form $\rho^2M_0(1$-${\bf S}^2_z)$.
Hence, although $\rho$ is small, the effect of mixing is enhanced by the
{\it resonance} of the singlet diffusion mode and the $S_z$=0 component of
the triplet. However, with stronger spin-orbit coupling the system is
driven out of the resonance due to the term $\zeta ({\bf S\cdot N)}$ in
(\ref{mdiff}). 

\begin{figure}[hbt]
\begin{center}
\leavevmode
\epsfig{file=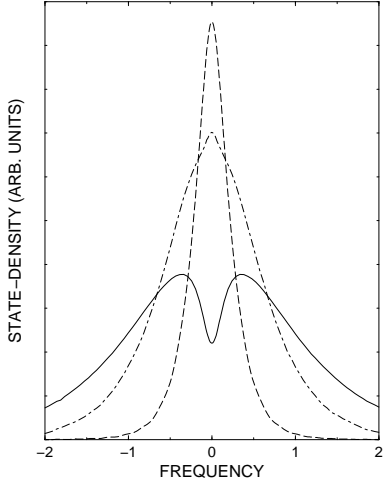,width=5.0cm,angle=0}
\end{center}
\caption{Fourier spectra of state-density for various Zeeman coupling
$\rho$ and SOI strength $\zeta$: $\rho$=0.006, $\zeta$=1 (solid curve);
$\rho$=0.0033, $\zeta$=1 (dot-dashed curve); $\rho$=0.006, $\zeta$=2.5
(dahsed curve). The frequency is defined as
$\omega$=$2\pi$-$\nu\Phi_0/\pi R^2$.}
\label{fig1}
\end{figure}

After substituting (\ref{expansion}) into (\ref{corr}) and carrying out
the integration, we need to perform a numerical summation over $n$ in order
to obtain the conductance fluctuations and the state-density fluctuations
which, according to Ref.~\onlinecite{shklovskii}, are given by the first
term of (\ref{corr}). The summation over $n$ is from -$n_{max}$ to
$n_{max}$=50, which gives converging results. The magnetic field $B_0$
is set at a value corresponding to a flux of 300$\Phi_0$. Within a
reasonable range of material parameters our numerical results depend only
weakly on the value of $B_0$.

\begin{figure}[hbt]
\begin{center}
\leavevmode
\epsfig{file=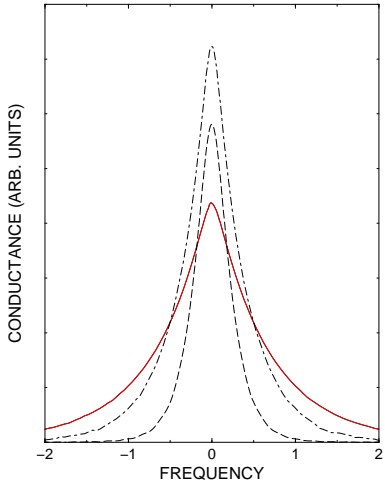,width=5.0cm,angle=0}
\end{center}
\caption{Fourier spectra of conductance: $\rho$=0.006, $\zeta$=1
(solid curve); $\rho$=0.0033, $\zeta$=1 (dot-dashed curve); $\rho$=0.006,
$\zeta$=2.5 (dahsed curve). The frequency is
$\omega$=$2\pi$-$\nu\Phi_0/\pi R^2$.}
\label{fig2}
\end{figure}  

Figs.~1-3 were calculated with $a$=$2\cdot10^{-4}$ and
$1/\tau_{\varphi}E_T=10^{-2}$. For the convenience of presentation, the
{\it frequency} in Figs.~1 and 2 is defined as
$\omega$=$2\pi$-$\nu\Phi_0/\pi R^2$ which is dimensionless. Fig. 1 shows the
main Fourier peak of the state density correlator. The splitting of the peak
(solid curve) for $\rho$=0.006 and $\zeta$=1 is due to the resonance of
diffusion modes. As the Zeeman coupling is reduced to $\rho$= 0.0033
(dot-dashed curve), or as the resonance is detuned by increasing the SOI to
$\zeta$=2.5 (dashed curve), the phenomenon of peak splitting disappears. We
have also found vanishing of this splitting when the value of the parameter
$a$ is enhanced, corresponding to a decrease of the magnetic field
correlation range of mesoscopic fluctuations. While the splitting is seen in
the state-density peak, it is absent in the diffusion coefficient spectrum.
In the parameter regime considered here, the contribution to the conductance
correlator from the diffusion coefficient spectrum is larger than that from
the state-denstiy correlator. Consequently, there is no peak splitting in
the Fourier spectrum of conductance oscillations as shown in Fig. 2.

\begin{figure}[hbt]
\begin{center}
\leavevmode
\epsfig{file=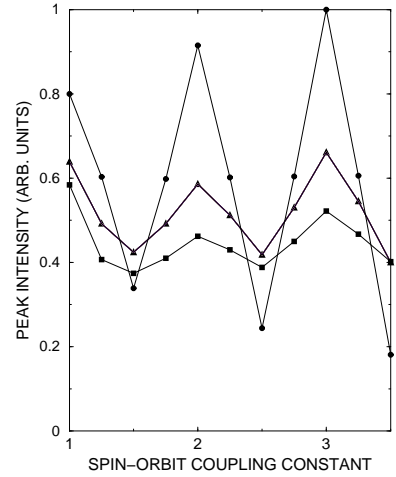,width=5.0cm,angle=0}
\end{center}
\caption{The intensity of the main Fourier peak as a function of the SOI
coupling: $\rho$=10$^{-4}$ (cicular dots); $\rho$=5$\cdot$10$^{-4}$
(triangles); $\rho$=10$^{-3}$ (squares).}
\label{fig3}
\end{figure} 

Our theory predicts a quantum oscillation of the intensity of the main peak
in the Fourier spectrum of conductance oscillation, as a function of the SOI
strength. To demonstrate that its origin lies in the geometric phase, let us
first set the Zeeman interaction $\rho$=0. In this case (\ref{mdiff}) can be
easily diagonalized by choosing the spin quantization axis along ${\bf N}$.
The eigenvalues of $\zeta {\bf S\cdot N}$ are $\zeta$, 0 and $-\zeta$.
Including the singlet state, the eigenvalue 0 is doubly degenerate.
The dependence of $\langle |g(\nu)|^2\rangle$ in (\ref{gnu}) from $\zeta$ 
can then be readily calculated,
because the $\pm\zeta$ can be absorbed by a shift of $\Delta\phi$ in
corresponding components of the correlator. If we ignore the small change
produced by this shift in the $a(\Delta\phi)^2$ term, after taking trace,
$\langle |g(\nu)|^2\rangle$ is found to be proportional to
1+$\cos[(2\pi$-$\omega)\zeta]$. At the center of the peak, $\omega$=0 and
so $\langle |g(\nu)|^2\rangle$ oscillates with $\zeta$ as $\cos^2\pi\zeta$.
Since the Zeeman interaction breaks the time inversion symmetry and thus
leads to an additional dephasing, the magnetic field correlation length is
reduced and hence the dependence on the geometric phase gets weaker. As a
result, the oscillating amplitude of $\langle |g(\nu)|^2\rangle$ decreases
rapidly with increasing Zeeman interaction. The numerical result of
this oscillation is shown in Fig.~3 for $\rho$=10$^{-4}$ (cicular dots),
$\rho$=5$\cdot$10$^{-4}$ (triangles), and $\rho$=10$^{-3}$ (squares).

The splitting of the main Fourier peak has been observed in a recent
experiment on AlSb/InAs/AlSb quantum well rings with radii about 1
$\mu$m~\cite{exp}. The material parameters for these samples are
$l$=10$^{-4}$cm, $g$=14, and $\alpha$$\simeq$10$^{-9}$~eV~cm~\cite{parsoi},
which give $\rho$$\simeq$10$^{-3}$ and $\zeta\simeq$3. With these values
of $\zeta$ and $\rho$, our calculation does not yield the splitting of the
main Fourier peak in both the state-density and the conductance. However,
this is not a total surprise because the samples used in
Ref.~\onlinecite{exp} are very small with $l$$\simeq$$R$, which is nearly
in ballistic regime instead of in diffusion regime. Furthermore, in these
samples the spin precession length $\hbar^2 /\alpha m^*$ is shorter than
the elastic mean free path, for which our perturbative treatment of SOI is
not valid.

In order to judge how relevant are our qualitative results to the above
mentioned experiment, let us consider the opposite limit of an ideal 1D
ring. The period of the Aharonov-Bohm (AB) oscillations changes due to the
dependence of the geometric phase and dynamic spin-phase on the magnetic
field. We will consider first the effect of geometric phase. In the region
of our interest $\rho'\phi$$\ll$$(\zeta^2$-$1)^{1/2}$ with $\rho'$ defined
as $\rho'$=$gm^*/k_fRm$, for an ideal ring one gets from
Refs.~\onlinecite{al,yqs} the geometric phase
$\theta_g$=$\pm\pi\rho' (\zeta^2$-$1)^{-1/2}\phi$+$C$ , where $C$ is a
constant independent of the magnetic field, and the $\pm$ signs refer to
the two electron spin orientations. Combining the geometric phase to the AB
phase $2\pi\phi$, we see that SOI leads to a split of the AB oscillations
in the transmittance of the ring into two oscillations with close
frequencies. However, this frequency splitting is about two orders of
magnitude less than the observed value~\cite{exp}. Next, we consider the
split of the AB oscillation frequency caused by dynamic spin phases, which
depend on the magnetic field in the form $\pm\pi\delta k_fR$, where
$\pm\delta k_f$ are the shifts of the Fermi wave-vector for up- and
down-spin electrons. The shift $\pi\delta k_fR$ has its maximum value
$\pi\rho'\phi/2$ in the absence of the SOI, and decreases with increasing
SOI strength~\cite{al,yqs}. Even at the largest value $\pi\rho'\phi/2$, the
corresponding split in AB oscillation frequency is of the order $\rho'$
which is too small to explain the experimental value.

The general feature of the peak splitting for the ideal 1D ballistic ring is
then similar to that for the diffusive 2D ring. Furthermore, in the 2D ring
the amount of the AB oscillations splitting as seen in the state-density
correlations also decreases with increasing SOI. If we use the same value
for $\rho'$ in a 1D ballistic ring and for $\rho$ in a 2D diffusive ring,
which means the same strength of the Zeeman coupling in both systems, in the
absence of SOI, the peak splitting in state density correlations is larger
than that in the transmittance of a 1D ring. This is due to the longer paths
traversed by a diffusing particle in its random walk along a ring, and so
acquiring a larger dynamic phase. We have reached the conclusion that the
two quite distinct limiting cases have led to the same qualitative picture.
Consequently, we conjecture that the peak splitting observed in near
ballistic 2D samples~\cite{exp} is due to the Zeeman interaction which is
not completely suppressed by the SOI and the mesoscopic fluctuations.

We acknowledge the support of the Royal Swedish Academy of Science under the
Research Cooperation Program between Sweden and the former Soviet Union,
Grant No. 12527, and of the Russian Foundation for Basic Research under
Grant No. 97-02-17324.


\begin{references}
\bibitem{comment} In fact, experimentally measured was the magnetoresistance. 
However, we are interested in the metallic regime. Hence, the
statistical properties of conductance and resistance are the same.
\end{references}
\end{document}